# From Simple Sensors to Complex Context: Insights for HabiTech

Albrecht Kurze
Chemnitz University of Technology
Germany
albrecht.kurze@informatik.tu-chemnitz.de

Karola Köpferl
Chemnitz University of Technology
Germany
karola.koepferl@hsw.tu-chemnitz.de

## ABSTRACT

We relate our previous as well as ongoing research in the domain of smart homes to the concept of HabiTech. HabiTech can benefit from existing approaches and findings in a broader context of whole buildings or communities within. Along with data comes context of data capture and data interpretation in different dimensions (spatial, temporal, social). For defining what is 'community' proximity plays a crucial role in context, both spatially as well as socially. A participatory approach for research in living in sensing environments is promising to address complexity as well as interests of different stakeholders. Often it is the complex context that makes even simple sensor data sensitive, i.e. in terms of privacy. When it comes to handle shared data then concepts from the physical world for shared spaces might be related back to the data domain.

## 1 INTRODUCTION

The Internet of Things has arrived in the 'smart' building – in office spaces as well as in homes. More and more sensors capture data – either pre-installed as part of the building infrastructure or brought to the premises as smart and connected sensor equipped (everyday) objects and appliances. All these sensors allow an omnipresent environmental monitoring. Potential problems come in as soon as data is captured and processed – as such, data is nearly never created, handled or used by a single person only. As soon as such sensor data is shared a whole bunch of implications unfolds. The situation tends to get even more serious as soon as spaces are shared and situated knowledge [9] (as context of data) is involved.

Based on the results of two finished projects (*Miteinander* and *Data-I*) and considerations for a new, recently launched project (*Simplications*), we briefly discuss the possible relevance of insights for HabiTech.

## 2 FROM DATA TO CONTEXT OF DATA

Besides the data itself as measured values it is important to understand the context of the data (the how) – technical well as non-technical. Important technical aspects are the technology itself, mainly the sensors and their technical capabilities etc. but also their parameterization to become "senseful sensors" [10] for a specific purpose and the following data processing. In brainstorming session based on pre-studies for collecting sensor data in different contexts we identified critical non-technical dimensions and characteristic therein for context of data from the building:

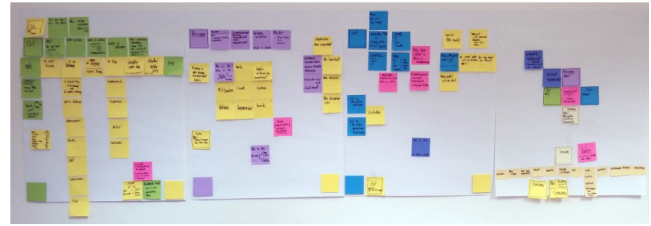

Figure 1: Results of brainstorming the dimensions of place, time, and persons for context of data capture and interpretation.

**Spatial - place - Where:** inside vs. outside, private vs. semi public vs. public, exclusive vs. shared - in the home, in the office, in the building in the public, in different rooms within a flat, in rooms with a specific function/meaning, e.g. kitchen, bathroom, bedroom.

**Temporal - time - When:** e.g. time, day, time of day, weekdays, weekend, how often etc.; short-, mid- and long-term; parallel and consecutive uses

**Social - persons - Who:** e.g. actors, stakeholders (primary, secondary, tertiary), roles, relations, by-standers (e.g. visitors or care-givers) etc.

While spatial might be intuitive, temporal is also important for space used or inhabited at different times and time-scales - throughout a day, a week or even longer time-spans of years, e.g. consecutive tenants in the same apartment. Would they share their smart environment data in the sense of passing it on to the next tenant, e.g. smart thermostat data, and if yes how?

Often a meaningful approach to characterize the context of data capture and interpretation cannot be limited to an isolated dimension. In our experience, it is important to think in intertwined combinations of these dimensions, especially for building data. Often a combination of 'where' and 'when' gives a hint for the 'who' of data, e.g. within a flat but also within an office building. When it comes to sensemaking of simple sensor data also additional dimensions of what and even why have to be considered [16].

Thinking in the dimensions mentioned above helped not to shaped our work in the research projects *Miteinander* and *Simplications* but also in the teaching-learning project *Data-I* for students in their work with sensor data from the home [13].

*Relevance for HabiTech:* The 'where', 'when' and 'who' are of general relevance especially to understand potential differences between shared and private spaces, e.g. for communities of inhabitants in buildings or neighborhoods.

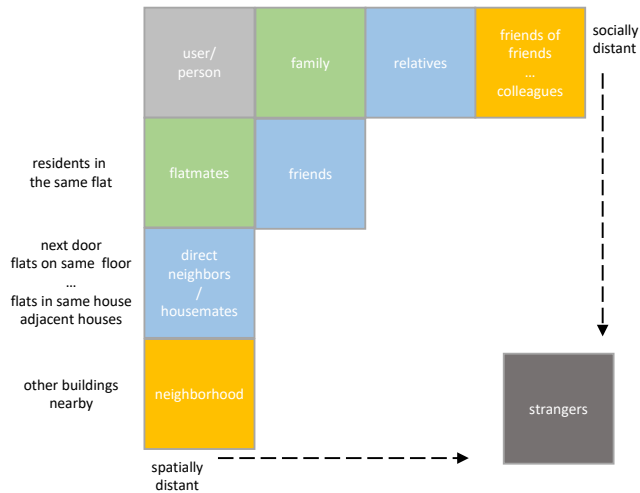

Figure 2: Example of concept 'proximity' for sharing data.

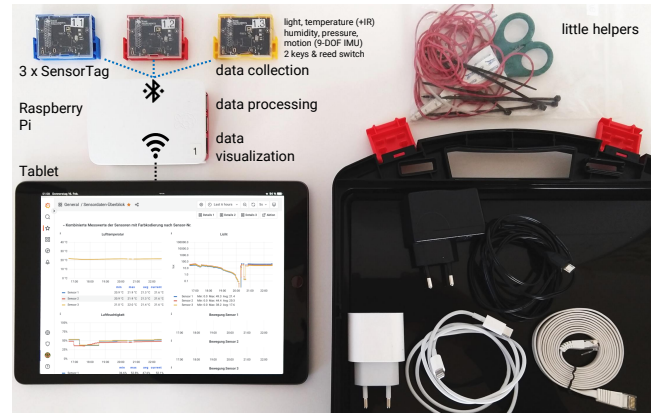

Figure 3: Components of the Sensorkit (sensors, Raspberry Pi, tablet pc, etc.) - 'all in a box' and ready for use 'out of the box'.

## 3 DEFINING 'COMMUNITY' IN CONTEXT OF DATA

Proximity seems to be a viable concept for defining communities, e.g. for sharing sensor data within 'close' proximity. The concept of proximity can built on multiple approaches but might need more than one dimension (figure 2). Is proximity defined by space, such as adjacent rooms within a flat, neighboring flats, flats in the same block or houses next to each other etc.? Or is it defined socially by relation of persons, such as family, friends, neighbors or colleagues etc.?

The context of data is not only important for meaningful capture of data but also for its interpretation [5, 6, 12, 15]. Closer proximity typically leads to potential for serious implications, e.g. for privacy and in terms of surveillance as the data can be contextualized and interpreted with additional "situated knowledge" [9] . Even a little bit of additional knowledge might turn a guess on data into an educated guess [12], e.g. when somebody left home or came back. A far distant 'stranger' such as a landlord might have by far less context and knowledge than those that are members of a building community that is sharing data and situated knowledge for context.

*Relevance for HabiTech:* Proximity has to be thought beyond a spatial dimension, especially for sharing data. We therefore encourage a holistic approach that takes into account not only technological aspects, but also social dynamics and ethical considerations in the design and use of buildings and concepts, and balances these with the needs of the occupants, so that self-determined and well-informed digital citizenship in residential buildings can be matched with the protection of privacy and autonomy in increasingly networked environments with high potential for use.

## 4 PARTICIPATORY RESEARCH FOR LIVING IN SENSING ENVIRONMENTS

For our participatory HCI research in the project *Miteinander* (2014-2022) we selected different appealing settings of context with a few basic constrains. We focused on 'simple' sensors, e.g. luxmeter, hygrometer, thermometer etc. (no cameras or microphones), mainly in 'the home' and for user studies of a few weeks.

We used our *Sensing Home Kit* [1, 3] in a number of user studies with variations in space (different flats and community living) and in social relation of the participating persons (relatives, neighbors, members of community living, etc.).

In our new project *Simplications* (2023-2026) we will built on the positive experience. Together with consumers, we will conduct participatory research for privacy by co-design on the possible implications of using seemingly simple but networked sensors (e.g. light, temperature, motion) in the home. Our findings will be used to develop and evaluate media and interventions for digital education that will contribute to the informed design and use of sensor data applications in the home.

For *Simplications* we extended the abilities and potential use cases of our sensing technology into the *Sensorkit*. While it continues the established focus on simple sensors it also integrates additional data from the building environment, in specific air quality data ($CO_2$ level and an air quality index based on volatile organic compounds and other air pollutants) as well as noise/loudness (sound pressure level fast).

Both mentioned projects in common is a participatory research approach that involves multiple formats engagement of those that will be influenced by the implementation and use of technology - a concept in research and development gaining grip and popularity [14]. While particiaption is not only relevant in the building - it is of extraordinary importance there - cause by complexity of context, i.e. when considering different roles and stakeholders there (see dimensions of 'Who' before).

The use and integrations of data from different sources (sensors, interviews, ...), in different characteristics (quantitative vs. qualitative), granularity/richness (thin vs. thick), size (small vs. big), and various roles (inspirational/explorative, constructional, evaluative) along the human-centered design process (HCD) has resulted in the concept of data as human-centered design material [7, 8]. A clear understanding

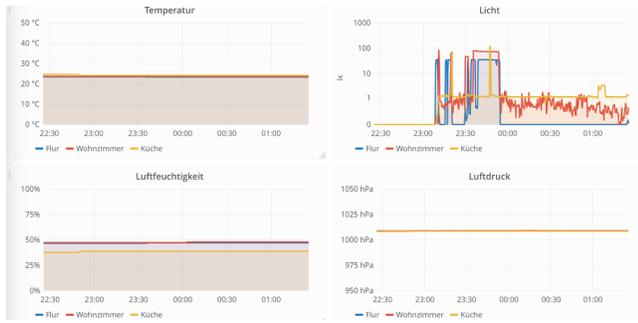

**Figure 4: Example of data visualization in Guess the Data.**

*Relevance for HabiTech:* For a successful use of data from the building for HabiTech these insights have to be considered. Research and development have to be done participatory and with data as human-centered design material.

## 5 FROM SIMPLE SENSORS TO SERIOUS IMPLICATIONS

For *Miteinander* we developed and field-tested the '*Guess the Data*' HCI method [12], which enabled people to use and make sense of live data from their homes and to collectively interpret and reflect on anonymized (raw) data from the homes (figure 4). In our study we focused on 'the home' but also how these homes are embedded in a broader context and contrasted them in different groups. We included a group of flats spread over the whole city, flats of relatives in the same neighborhood as well as flats in the same apartment block complex.

Our findings show how participants reconstruct behavior, both individually and collectively. We also found the expose the sensitive personal data of others, and use sensor data as evidence and for lateral surveillance within (multi-habitant) households. While in some scenarios such a use of sensor data was intentional we also found in majority of critical uses that a good intention in collecting the sensor data turned "accidentally evil" when making sense and use of it [2]. Good intentions and positive values such as efficiency, comfort and security are often in conflict with values such as privacy, autonomy and transparency - requiring careful consideration and justification. We were also able to show that close proximity - socially by relatives but also spatially by neighbors – gives additional situated knowledge for interpretation. Often this leads to very contra-intuitive wicked implications for privacy, e.g. that the sharing of simple sensor data might within close proximity, e.g. with trusted household or family members, might be more sensitive than the use by a seemingly almighty but distant 3rd party 'Big Brother' [11].

*Relevance for HabiTech:* These findings may be relevant to HabiTech and multi-occupancy buildings and their communities, as these contexts are not fundamentally different, but slightly broader, emphasising the importance of balancing individual privacy with community engagement. The Sensorstation concept serves as a practical approach to facilitate collaborative exploration of IoT applications in buildings, thereby enabling shared experiences and responsibilities without compromising on personal data security.

## 6 FROM SHARED SPACE TO SHARED DATA

We developed 'Sensorstation' [4] (figure 5), a research product that allows residents of a living community to explore the use of simple IoT sensor data. It uses simple IoT sensors in combination with a stationary device that consists of an input screen for configuration and an output screen for displaying notifications. The concept of Sensorstation introduces an abstraction layer to the raw data. In a configuration interface, users can set conditions, e.g. events, thresholds or range, for the individual color-coded sensors and sensor functions (motion, light, temperature, humidity, air pressure) and assign a short message to this condition. Every time a condition is met a notification is sent to a shared message stream that is displayed on the output screen. The stationary part is intended to be placed in a central position in the flat, accessible and readable for all inhabitants, e.g. on the kitchen table, while the sensors can be used everywhere in the flat.

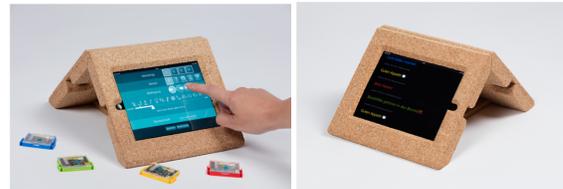

**Figure 5:** *Sensorstation* for shared data capture and access a) input screen side and sensors, b) output screen side with shared message stream for all residents.

Our findings show that the participants explored a number of different smart home applications with Sensorstation. These ranged from sensor applications for creating positive connections within the group, over self-monitoring, to control over others as well as reward systems and penalties. Sensorstation in combination with the group discussion not only provided insight into the use of sensor data in the shared apartment, but also into the underlying usage motives. The influence of the technology on the group and possible chances and conflict potentials in the use of sensor technology lies in the implications of shared sensor data in shared spaces.

*Relevance for HabiTech:* Design for HabiTech has the obligation to consider smart technology that acknowledges boundaries between different types of space and data usage – such as walls, doors and curtains do in the real world.

## 7 SUMMARY AND CONCLUSION

In this article we summarized some of our insights from finished and ongoing projects in 'the home' involving simple sensor data from this context. We explained how these insights have relevance not only for the domain of the home but also in broader meaning for HabiTech: relevant dimension of (non-technical) context (spatial, temporal, social), how proximity has to be thought multi-dimensional, how participatory research can help to deal with complex contexts, a lens on privacy as an example how complex context plays a role in easily wicked implications, and how to think shared space in realms of shared data.

## ACKNOWLEDGMENTS

This research is funded by the German Ministry of Education and Research (BMBF) grant FKZ 16KIS1868K.

**Albrecht Kurze:** Albrecht Kurze is a post-doctoral researcher at the chair Media Informatics at TU Chemnitz and one of the principal investigators in Simplications. With a background in computer science his research interests are on the intersection of Ubiquitous HCI and human centered IoT: How do sensors, data and connectedness in smart products and environments allow for new interactions and innovation and how do we cope with the implications that they create, i.e. for privacy in the home.

**Karola Köpferl:** Karola Köpferl is a PhD candidate at TU Chemnitz in Simplications holding a Master degree in Care, social and health management. Karola's doctoral research focuses on how, why and with what attitudes and experiences older people use smart technologies. She is interested in the technical realization of all devices and sensors, as well as the practices of use and, above all, the awareness or lack of awareness of the implications of supposedly well-intentioned technological gifts and miracles.